\documentclass[11pt]{article}
\usepackage{amssymb}
\usepackage{amsmath}
\usepackage{graphics}
\usepackage{epsfig}
\usepackage{a4wide}
\usepackage{cite}
\usepackage{multirow}
\usepackage{color}
\usepackage{mathrsfs}

\DeclareGraphicsExtensions{.eps}

\begin{document}

\title{  NLO QCD + EW corrections to $ZZZ$ production with subsequent leptonic decays at LHC }
\author{ Wang Hong$^a$, Zhang Ren-You$^a$, Ma Wen-Gan$^a$, Guo Lei$^b$, Li Xiao-Zhou$^a$, and Wang Shao-Ming$^b$ \\
{\small $^a$ Department of Modern Physics, University of Science and Technology of China, }  \\
{\small $~~$ Hefei, Anhui 230026, P.R.China} \\
{\small $^b$ Department of Physics, Chongqing University, Chongqing 401331, P.R. China} }

\date{}
\maketitle \vskip 15mm
\begin{abstract}
In this paper we present the NLO QCD + NLO EW corrections to the $ZZZ$ production with subsequent $Z$-boson leptonic decays at the LHC by adopting the improved narrow width approximation, which takes into account off-shell contributions and spin correlations. Integrated cross sections at $13$, $14$, $33$ and $100~{\rm TeV}$ hadron colliders and various kinematic distributions are presented. Our numerical results show that the jet emission correction accounts for a large part of the total QCD correction, especially in the high energy region. By applying a proper cut to the jet transverse momentum, e.g., $p_{{\rm \,T,jet}} > p_{{\rm \,T,jet}}^{{\rm \,\,\,cut}}=50~{\rm GeV}$, the jet emission correction can be reduced and the cross section is less dependent on the factorization/renormalization scale. This work reveals that both the NLO QCD and the NLO EW corrections are significant. For example, the NLO QCD, NLO EW and NLO QCD + EW relative corrections in the inclusive event selection scheme at the $13~{\rm TeV}$ LHC can reach $63.2\%$, $-9.6\%$ and $47.5\%$, respectively, which means that neither the NLO QCD nor the NLO EW correction is negligible in precision calculations.
\end{abstract}


\vfill \eject \baselineskip=0.32in

\renewcommand{\theequation}{\arabic{section}.\arabic{equation}}
\renewcommand{\thesection}{\arabic{section}}

\makeatletter      
\@addtoreset{equation}{section}
\makeatother       

\par
\vskip 5mm
\section{Introduction}
\label{sec:intro}
\par
The discovery of the $126~{\rm GeV}$ Higgs boson at the Large Hadron Collider (LHC) in July 2012  \cite{Higgs-ATLAS,Higgs-CMS} has cuased tremendous progress to be made in particle physics and represents a great victory for the standard model (SM). One of the main tasks of future experiments is to determine the gauge couplings in the SM and test the validity of the gauge principle. Triple gauge boson production at the LHC provides an opportunity for the precision measurement of quartic gauge boson couplings, and it would help us to better understand the electroweak symmety breaking \cite{qgbc,ewsb}. In the past years, the theoretical predictions for all this kind of processes at the LHC have been calculated up to the QCD next-to-leading order (NLO) \cite{WWZ, ZZZ, WWW-WZZ, WWZ+ZZZ+WWW+WZZ, WWr-ZZr, Zrr-rrr, Wrr, Wrr-decay, WZr}, while the predictions up to the QCD + EW NLO for the $pp \to WWZ$ \cite{EWWWZ}, $pp \to WZZ$ \cite{EWWZZ} and $pp \to WWW$ \cite{EWWWW} processes are provided. Since precision measurements will be possible in the LHC Run2 or future colliders, the calculations of $VV'V''~ (V,V',V''=W~ {\rm or}~ Z)$ productions, including the subsequent vector boson decays at the QCD + EW NLO, are desired. This issue was listed in the Les Houches 2013 high precision wishlist \cite{wishlist-2013}, and still exists in the report of Les Houches 2015 \cite{wishlist-2015}.

\par
The $ZZZ$ production at the LHC provides a window for studying the quartic $ZZZZ$ coupling, and it also serves as a background to the supersymmetric tri-lepton production signature. In view of these reasons, precision calculations for this process are necessary in order to meet the requirements of the experimental measurement. The NLO QCD correction to $pp \to ZZZ$ without subsequent $Z$-boson decay was calculated in Refs.\cite{ZZZ} and \cite{WWZ+ZZZ+WWW+WZZ}, where the contributions from the diagrams with internal Higgs-boson exchange are neglected.

\par
In this work, we calculate the NLO QCD + NLO EW corrections to the $pp \to ZZZ + X$ process including $Z$-boson leptonic decays in the SM, and find that the NLO EW correction is considerable and cannot be ignored, even though it is suppressed by the smallness of the EW coupling constant $\alpha$. The paper is organized as follows: in section \ref{sec:calculation}, we report the calculation details, in section \ref{sec:num} we provide the integrated cross sections and the relevant kinematic distributions in the different event selection schemes, and finally, we provide a short summary.

\vskip 5mm
\section{ Calculation strategy }
\label{sec:calculation}
\par
\begin{figure}[ht!]
\centering
\includegraphics[scale=1]{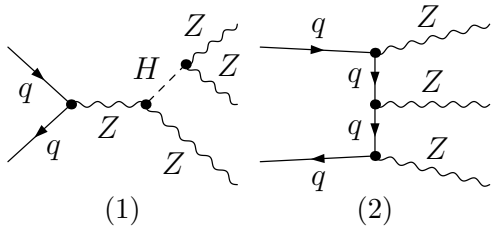}
\caption{\small The representative LO diagrams for the partonic process $q\bar{q}\to ZZZ$.}
\label{fig:lo}
\end{figure}

At the leading-order (LO) only the partonic channel $q\bar{q} \to ZZZ$ contributes to triple $Z$-boson production at the LHC. We adopt the five-flavor scheme and neglect the masses of $u$-, $d$-, $c$-, $s$- and $b$-quark throughout our calculations. The representative tree-level diagrams for the $q\bar{q} \to ZZZ$ subprocess are shown in Fig.\ref{fig:lo}. It should be noted that the Feynman diagrams involving internal Higgs-boson exchange, e.g., Fig.\ref{fig:lo}(1), are included in our calculations, even though their contributions are relative small, accounting for about $8\%$ at the LO.

\par
The NLO QCD calculation for the $pp \to ZZZ+X$ process was previously performed in Ref.\cite{ZZZ}, where the authors ignored the contributions from the diagrams with internal Higgs-boson exchange. In order to verify the correctness of our program, we have checked the NLO QCD corrected integrated cross section numerically by using both the {\sc MadGraph5} \cite{madgraph} package and our own program, with the same input parameters and considerations as those used in Ref.\cite{ZZZ}. The numerical results for the NLO QCD corrected integrated cross sections are listed in Table \ref{tab:cross-check}. We find that the results obtained from both packages are highly coincident with those in Ref.\cite{ZZZ}. Since the detailed description of the NLO QCD calculation has already been given in a previous paper \cite{ZZZ}, we provide only the calculation setup of the NLO EW corrections in this section.
\begin{table}[ht!]
\center
\begin{tabular}{cccc}
\hline
\hline
\\ [-1.7ex]
{}  &  ~~~~~~~~{Ours}~~~~~~~~   & ~{{\sc MadGraph5}}~ &  ~{Reference \cite{ZZZ}}~   \\
\\ [-1.7ex]
\hline
\\ [-1.7ex]
~$\sigma_{{\rm QCD}}$~(fb)~ & ~~~15.20(2)~~~ & ~~15.25(2)~~ & ~~~~~15.2~~~~~ \\
\\ [-1.7ex]
\hline
\hline
\end{tabular}
\caption{\small The NLO QCD corrected integrated cross sections obtained from our program, {\sc MadGraph5} and Ref.\cite{ZZZ} with $\mu_R = \mu_F = 3M_Z$.}
\label{tab:cross-check}
\end{table}

\par
In our calculation we adopt the 't Hooft-Feynman gauge. The Feynman diagrams concerned are automatically generated using {\sc FeynArts-3.7} package \cite{feynarts}, and the corresponding amplitudes are algebraically simplified by employing the {\sc FormCalc-7.3} program \cite{formcalc}. In order to deal with the numerical instabilities which appear in the evaluation of the scalar and tensor one-loop integrals, we use the improved {\sc LoopTools-2.8} package \cite{ff, formcalc} which can automatically switch to quadruple precision codes when the Gram determinant is sufficiently small \cite{gram-fb,gram-cc}.

\par
The NLO EW correction to the $pp \to ZZZ + X$ process includes the following four parts: (1) the EW virtual one-loop correction to $pp \to q\bar{q} \to ZZZ$; (2) the real photon emission correction from $pp \to q\bar{q} \to ZZZ + \gamma$; (3) the contribution of the collinear photon emission part of the quark parton distribution function (PDF) EW counterterm; (4) the photon-induced correction consisting of the contributions from $pp \to q/(\bar{q})\gamma \to ZZZ + q/(\bar{q})$ (photon-induced subprocesses) and the collinear quark emission part of the quark PDF EW counterterm.

\par
To analyze the origin of the NLO EW correction clearly, we separate the NLO EW correction into photonic (QED) and genuine weak corrections. The weak correction comes from the weak virtual one-loop diagrams without virtual photon exchange. The rest of the NLO EW corrections are called QED corrections. Then, the full NLO EW correction can be expressed as
\begin{eqnarray}
\label{full EW correction}
\Delta \sigma_{{\rm EW}}
&=&
\Delta \sigma^{q\bar{q},V}_{{\rm EW}} + \Delta \sigma^{q\bar{q},R}_{{\rm EW}} + \Delta \sigma^{q\bar{q},{\rm PDF}}_{{\rm EW}}+ \Delta \sigma^{q(\bar{q})\gamma}_{{\rm EW}} \nonumber \\
&=&
\Delta \sigma^{q\bar{q},V}_{{\rm W}}
+
\left(
\Delta \sigma^{q\bar{q},V}_{{\rm QED}} + \Delta \sigma^{q\bar{q},R}_{{\rm EW}} + \Delta \sigma^{q\bar{q},{\rm PDF}}_{{\rm EW}}+ \Delta \sigma^{q(\bar{q})\gamma}_{{\rm EW}}
\right) \nonumber \\
&=&
\Delta \sigma_{{\rm W}} + \Delta \sigma_{{\rm QED}} \,.
\end{eqnarray}

\par
In both the QCD and EW calculations, we isolate the ultraviolet (UV) and infrared (IR) divergences by adopting the dimensional regularization scheme in $d = 4-2\epsilon$ dimensions. We renormalize the relevant masses and fields in the on-mass-shell renormalization scheme. The definitions and expressions for the relevant renormalization constants can be found in Ref.\cite{Denner:EWphysics}.

\par
The renormalized electric charge is given by $e_{0} = ( 1 + \delta{Z_e} ) e$, where $e_0$ is the bare electric charge and $\delta Z_e$ is the corresponding renormalization constant. There are three schemes for the renormalization of the fine structure constant:

\par
(1) The $\alpha(0)$-scheme, where $\alpha = \alpha(0)$ is defined in the Thomson limit, and the electric charge renormalization constant $\delta Z_e$ is written as \cite{Denner:EWphysics}
\begin{eqnarray}\label{Log-1}
\delta Z^{\alpha(0)}_e = -\frac{1}{2}\delta Z_{AA} - \frac{1}{2} \tan\theta_{\rm W} \delta Z_{ZA} =
\frac{1}{2}\frac{\partial \sum^{AA}_{\rm T}(p^2) }{\partial p^2}\bigg{|}_{p^2\to 0} - \tan\theta_{\rm W} \frac{\sum^{AZ}_{\rm T}(0)}{M^2_Z},
\end{eqnarray}
where $\theta_{\rm W}$ is the weak mixing angle, and $\sum^{ab}_{\rm T}(p^2)$ denotes the transverse part of the unrenormalized self-energy at four-momentum squared $p^2$.
\par
(2) The $\alpha(M_Z)$-scheme, where $\alpha$ is evolved from zero four-momentum squared to the $Z$ pole by the renormalization-group equations.
\par
(3) The $G_{\mu}$-scheme, in which $\alpha_{G_{\mu}}$ is given by
\begin{eqnarray}
\label{alpha-G}
\alpha_{G_\mu} = \frac{\sqrt{2}}{\pi}G_{\mu}M_W^2\sin^2\theta_{\rm W} ,
\end{eqnarray}
and the corresponding electric charge renormalization constant is
\begin{eqnarray}\label{Log-4}
\delta Z^{G_{\mu}}_e = \delta Z^{\alpha(0)}_e - \frac{1}{2}\Delta r ,
\end{eqnarray}
where $G_{\mu}$ is the Fermi constant, and $\Delta r$ is the NLO EW correction to the muon decay. The explicit expression for $\Delta r$ can be written as \cite{w-production, delt-r}
\begin{eqnarray}\label{Log-3}
\Delta r &=& -\delta Z_{AA} - \cot^2\theta_{\rm W} \left( \frac{\sum^{ZZ}_{\rm T}(M^2_Z)}{M^2_Z} - \frac{\sum^{WW}_{\rm T}(M^2_W)}{M^2_W} \right) + \frac{\sum^{WW}_{\rm T}(0)-\sum^{WW}_{\rm T}(M^2_W)}{M^2_W} \nonumber \\
&& +\, 2 \cot\theta_{\rm W} \frac{\sum^{AZ}_{\rm T}(0)}{M^2_Z} +
\frac{\alpha(0)}{4\pi \sin^2\theta_{\rm W}} \left( 6 + \frac{7 -4
\sin^2\theta_{\rm W}}{2 \sin^2\theta_{\rm W}} \ln(\cos^2\theta_{\rm W}) \right).
\label{eqn:deltar}
\end{eqnarray}

\par
For the process without an external photon line at the LO, in both the $\alpha(M_Z)$-scheme and the $G_{\mu}$-scheme the large light-fermion logarithms resulting from charge renormalization in the EW correction can be cancelled. Furthermore, according to the discussions in Refs.\cite{wishlist-2013} and \cite{Denner:EWphysics}, it is suggested to use the $G_{\mu}$-scheme, then one can obtain some significant universal corrections to the LO contributions connected to the renormalization of the weak mixing angle.

\par
In calculating the real photon emission and the photon-induced partonic processes, we use the two cutoff phase space slicing (TCPSS) method \cite{TCPSS} to isolate the soft and collinear IR singularities. The singularities originating from these processes are partially cancelled by the photonic IR singularities from an exchange of the virtual photon in the loop, and the remaining singularities are absorbed by the related quark PDF EW counterterms. Then we get IR-safe results and compare the integrated cross section results with those obtained by using the dipole subtraction (DS) method \cite{DIPOLE-1996, DIPOLE-1997, DIPOLE-2000, DIPOLE-alpha-1, DIPOLE-alpha-2}. We find that they are perfectly coincident with each other. In further calculations, we use the DS method to calculate the integrated cross section in order to get smaller Monte Carlo errors, and adopt the TCPSS method in evaluating the kinematic distributions to avoid the so-called missed binning problem in the DS method \cite{DIPOLE-alpha-2}. We take the cutoffs in the TCPSS method as $\delta_s = 10^{-4}$ and $\delta_c = 10^{-5}$, and set $\alpha^{{\rm DS}} = 0.1$ in the DS method \cite{DIPOLE-alpha-1} to control the volume of the dipole phase space.

\par
The quark PDF EW counterterm $\delta \Phi_{q|P}^{{\rm EW}}$ consists of the collinear photon emission term ($\delta \Phi_{q|P}^{{\rm EW},(\gamma)}$) and collinear light-quark emission term ($\delta \Phi_{q|P}^{{\rm EW},(q)}$). These two collinear terms are expressed as below \cite{EWWWZ,DIS-facscheme} in the DIS factorization scheme:
\begin{eqnarray}
\label{PDF-EWCT}
&& \delta \Phi_{q|P}^{{\rm EW},(\gamma)}(x,\mu_F,\mu_R) =
   \frac{Q_q^2\alpha}{2 \pi}\int_x^1 \frac{dz}{z}\Phi_{q|P}(x/z,\mu_F) \nonumber \\
&&  ~~~~~~~~~~~~~~~~~~~~~~~~~~~~~~ \times
   \left\{
   \frac{1}{\epsilon} \frac{\Gamma(1 - \epsilon)}{\Gamma(1 - 2 \epsilon)}\left( \frac{4 \pi \mu_R^2}{\mu_F^2} \right)^{\epsilon} \left[ P_{qq}(z) \right]_+ - C_{qq}^{{\rm DIS}}(z) \right\}, ~~~~~~~~\nonumber \\
&& \delta \Phi_{q|P}^{{\rm EW},(q)}(x,\mu_F,\mu_R) =
   \frac{3Q_q^2\alpha}{2 \pi} \int_x^1 \frac{dz}{z} \Phi_{\gamma|P}(x/z,\mu_F) \nonumber \\
&&  ~~~~~~~~~~~~~~~~~~~~~~~~~~~~~~ \times
   \left\{ \frac{1}{\epsilon} \frac{\Gamma(1 - \epsilon)}{\Gamma(1 - 2 \epsilon)}\left( \frac{4 \pi \mu_R^2}{\mu_F^2} \right)^{\epsilon} P_{q\gamma}(z) - C_{q\gamma}^{{\rm DIS}}(z)  \right\},
\end{eqnarray}
where $\mu_F$ and $\mu_R$ are the factorization scale and renormalization scale, respectively. $Q_q$ is the electric charge carried by the initial quark $q$. The splitting functions $P_{qq}(z)$ and $P_{q\gamma}(z)$ are given by
\begin{eqnarray}
\label{eq:spliting}
P_{qq}(z) = \frac{1 + z^2}{1 - z}, ~~~~~ P_{q\gamma}(z) = z^2 + (1 - z)^2,
\end{eqnarray}
and the $\left[ \ldots \right]_+$ prescription is understood as
\begin{eqnarray}
\int_x^1 dz \left[ g(z) \right]_+ f(z) = \int_x^1 dz \, g(z) \left[ f(z) - f(1) \right] - f(1) \int_0^x dz \, g(z) \,.
\end{eqnarray}
The coefficient functions for the DIS scheme are given by \cite{EWWWZ,DIS-facscheme}
\begin{eqnarray}
&& C_{qq}^{{\rm DIS}}(z)
   =
   \left[
   P_{qq}(z) \left( \ln \frac{1-z}{z} - \frac{3}{4} \right) + \frac{9 + 5 z}{4} \right]_+ , ~~~~~~~~~\nonumber \\
&& C_{q\gamma}^{{\rm DIS}}(z)
   =
   P_{q\gamma}(z)
   \ln \frac{1-z}{z} - 8z^2 + 8z - 1 \,.
\end{eqnarray}

\par
The emission of the photon collinear to the decayed lepton leads to a correction enhancement due to large logarithms, and influences the kinematic distributions and acceptance rate of the leptons. However, large logarithms from the final-state radiation (FSR) could be prevented by defining ``dressed" leptons \cite{FSR,dress-lepton}, which would treat the collinear lepton-photon system as one quasi-particle. Usually, the final-state electrons are detected automatically as ``dressed" electrons by an electromagnetic calorimeter, while muons are detected at the LHC as ``bare" particles by other detectors, without an automatic combination with the photons. A similar situation happens with tau leptons. In order to reduce large FSR correction, the observed ``bare" leptons can be reconstructed as ``dressed" leptons via photon recombination \cite{wishlist-2013}. Thus, in the NLO EW calculation, we neglect the corrections from the real photon radiation off the final charged leptons, and consider the final charged leptons as ``dressed" leptons in discussing their kinematic distributions.

\vskip 5mm
\section{Numerical results and discussion}
\label{sec:num}
\par
\subsection{Input parameters}
\label{sec:inputpara}
\par
The relevant SM input parameters in our calculation are listed as follows \cite{PDG}:
\begin{eqnarray}
\label{eq:SMpar}
&& M_W = 80.385~{\rm GeV},~~ M_Z = 91.1876~{\rm GeV},~~ M_H = 125.09~{\rm GeV}, \nonumber \\
&& M_t=173.21~{\rm GeV},~~ G_{\mu} = 1.1663787 \times 10^{-5}~{\rm GeV}^{-2} \, .  ~~~~~~
\end{eqnarray}
The $Z$-boson decay width of $\Gamma_{\textrm{total}}^Z=2.4439~{\rm GeV}$ in the fixed width scheme is obtained by using the {\sc MadSpin} program with the above input parameters. We set all the leptons and light quarks, including the bottom quark, to be massless. Considering no charged boson exchange in the LO graphs for the $ZZZ$ production and the unitarity of the Cabibbo-Kobayashi-Maskawa (CKM) matrix, it is acceptable to take CKM matrix as a unit matrix. For simplicity we take the renormalization and factorization scales as being equal ($\mu_R = \mu_F = \mu_0 = \frac{3}{2} M_Z$) in case there is no other statement, and $\mu_0$ is defined as the central scale.

\par
The EW corrections from the photon-induced partonic processes are convoluted with the photon distribution function. The NNPDF2.3QED PDFs \cite{NNPDF} are ideal options for this requirement. The value of the strong coupling constant is quoted as $\alpha_s(M_Z) = 0.119$ from the PDF set with five active flavors. The $\overline{MS}$ factorization scheme is used throughout the NLO QCD calculation, while the DIS factorization scheme is applied in the NLO EW calculation \cite{DIS-facscheme}.

\par
\subsection{Total cross sections}
\label{sec:totalcs}
\par
In analogy to the NLO EW correction, the full NLO QCD correction to the $pp \rightarrow ZZZ + X$ process can be expressed as
\begin{eqnarray}
\label{full QCD correction} \Delta \sigma_{{\rm QCD}} = \Delta
\sigma^{q\bar{q},V}_{{\rm QCD}} + \Delta \sigma^{q\bar{q},R}_{{\rm QCD}} + \Delta
\sigma^{q\bar{q},{\rm PDF}}_{{\rm QCD}}+ \Delta \sigma^{q(\bar{q})g}_{{\rm QCD}} \,.
\end{eqnarray}
The NLO QCD and EW relative corrections are given by
\begin{eqnarray}
\label{relative correction}
\delta_{{\rm QCD}} = \frac{\Delta \sigma_{{\rm QCD}} + \left( \sigma_{0} - \sigma_{{\rm LO}}
\right)}{\sigma_{{\rm LO}}},~~~~~~~~ \delta_{{\rm EW}} = \frac{\Delta
\sigma_{{\rm EW}}}{\sigma_{0}} = \frac{\Delta \sigma_{{\rm W}} + \Delta \sigma_{{\rm QED}}}{\sigma_{0}} = \delta_{{\rm W}} + \delta_{{\rm QED}} \,,
\end{eqnarray}
where $\delta_{{\rm W}}$ and $ \delta_{{\rm QED}}$ are the weak and QED relative corrections, respectively. In the above definitions, $\Delta \sigma_{{\rm QCD}}$ and $\Delta \sigma_{{\rm EW}}$ are evaluated by using NLO PDFs, and $\sigma_{{\rm LO}}$ and $\sigma_{0}$ are LO cross sections calculated using the LO and NLO PDFs, respectively. The expression $\delta_{{\rm QCD}}$ in Eq.(\ref{relative correction}) indicates that $\delta_{{\rm QCD}}$ contains the NLO QCD contributions from both the dynamic matrix element and the PDFs. On the other hand, the EW correction is normalized by $\sigma_{0}$ but not $\sigma_{{\rm LO}}$, which guarantees $\delta_{{\rm EW}}$ to be free from the QCD correction effects in the NLO PDFs, and exposes the matrix element correction effects in a more transparent way. The full NLO QCD + EW corrected cross section is obtained by combining the NLO QCD and EW corrections via the following naive product \cite{Denner-delta}:
\begin{eqnarray}
\label{Xection-nlo}
\sigma_{{\rm NLO}} &=& \sigma_{{\rm LO}} \left( 1 + \delta_{{\rm NLO}} \right) \nonumber \\
&=& \sigma_{{\rm LO}} \left( 1 + \delta_{{\rm QCD}} \right) \left( 1 + \delta_{{\rm EW}} \right).
\end{eqnarray}

\par
The numerical results of the LO, NLO QCD, NLO QCD + EW corrected integrated cross sections and the corresponding relative corrections for the $ZZZ$ production at the $13$, $14$, $33$ and $100~ {\rm TeV}$ hadron colliders are listed in Table \ref{tab:cross-sections}. The results are compared in the table by using different event selection schemes. We call the inclusive scheme without any cuts scheme I, and the exclusive scheme with the jet veto condition of $p_{{\rm \,T,jet}} > 50~{\rm GeV}$ scheme II. From the table, we can see that the NLO QCD relative correction is sizable, and increases steadily from $63.2\%$ at the $13~ {\rm TeV}$ to $91.4\%$ at the $100~ {\rm TeV}$ in scheme I. However, it decreases sharply to be no more than $24\%$ when the jet veto cut of $p_{{\rm \,T,jet}} > 50~{\rm GeV}$ is applied. This means that the real gluon and light-quark emission contributions are significant and account for a large part of the full NLO QCD correction.

\par
On the other hand, although suppressed by the smallness of the EW fine structure constant, the NLO EW correction is surprisingly remarkable in all chosen colliding energies and event selection schemes. Differing from the dramatic reduction in the NLO QCD relative correction by applying a jet veto, the NLO EW relative correction is pretty stable at about $-10\%$. The QED relative correction $\delta_{{\rm QED}}$ is listed in Table \ref{tab:cross-sections} as well. We can see that $\delta_{{\rm QED}}$ is about $-1.3\%$ for all the colliding energies and event selection schemes considered. Thus, we can conclude that the NLO EW correction mainly comes from the weak correction.

\par
In our calculation, we find that the NLO contribution from the initial photon-induced partonic processes is only about $0.1\%$ of the full NLO EW correction, i.e., $\left|\frac{\Delta \sigma^{q\gamma}_{{\rm EW}}}{\Delta \sigma_{{\rm EW}}}\right| \sim 0.1\%$, therefore the theoretical uncertainty from the photon PDF can be neglected.
\begin{table}[ht!]
\center\small
\begin{tabular}{ccccccccc}
\hline
\hline
\\ [-1.7ex]
{$\sqrt{s}$~[{\rm TeV}]}  &  {Scheme}   & ~{$\sigma_{{\rm LO}}$~(fb)}~ &  ~{$\sigma_{{\rm QCD}}$~(fb)}~  &  ~{$\sigma_{{\rm NLO}}$~(fb)}~ & {$\delta_{{\rm QCD}}$~(\%)} & {$\delta_{{\rm QED}}$~(\%)} & {$\delta_{{\rm EW}}$~(\%)} &{$\delta_{{\rm NLO}}$~(\%)}    \\
\\ [-1.7ex]
\hline
\\ [-1.7ex]
\multirow{2}{*}{13}  & I & 9.002(2) & 14.688(3) & 13.277(4) & 63.17 & -1.30 & -9.61 & 47.50 \\
&  II & 9.002(2) &  11.140(3) &  10.068(3) & 23.76 & -1.32 & -9.63  & 11.85 \\
\\ [-1.7ex]
\\ [-1.7ex]
\multirow{2}{*}{14}  & I & 10.118(2) & 16.599(3) & 14.994(4) & 64.06 & -1.29 & -9.67 & 48.20 \\
&  II & 10.118(2)  &  12.450(3) &  11.244(4) & 23.05 & -1.31 & -9.69 & 11.13 \\
\\ [-1.7ex]
\\ [-1.7ex]
\multirow{2}{*}{33}  & I & 34.162(7) & 60.15(2) & 54.01(2) & 76.07 & -1.29 & -10.21 & 58.09 \\
&  II & 34.162(7)  & 39.55(2) & 35.50(2) &  15.77 & -1.31 & -10.23 & 3.93 \\
\\ [-1.7ex]
\\ [-1.7ex]
\multirow{2}{*}{100} & I & 131.46(3) & 251.65(8) & 224.85(8) & 91.43 & -1.34 & -10.65 & 71.05 \\
& II & 131.46(3) &  140.43(8) & 125.45(9) & 6.83 & -1.36 & -10.67  & -4.57 \\
\\ [-1.7ex]
\hline
\hline
\end{tabular}
\caption{\small The LO, NLO QCD, NLO QCD + EW corrected integrated cross sections and the corresponding relative corrections for the $ZZZ$ production at the $13$, $14$, $33$ and $100~ {\rm TeV}$ hadron colliders in scheme I and II, separately.  }
\label{tab:cross-sections}
\end{table}

\par
We provide the dependence on the factorization and renormalization scales of each corrected cross section at the $13~{\rm TeV}$ LHC in Table~\ref{tab:mu-dependence}. For simplicity, we set $\mu_F = \mu_R = \mu$ in our calculation. The scale uncertainty is defined as
\begin{eqnarray}
\eta = \frac{\max\Big\{\sigma(\mu) | \frac{1}{4} \mu_0 \leq \mu
\leq 4 \mu_0 \Big\} - \min\Big\{\sigma(\mu) | \frac{1}{4} \mu_0
\leq \mu \leq 4 \mu_0 \Big\}}{\sigma(\mu_0)}.
\end{eqnarray}
From Table~\ref{tab:mu-dependence} we can figure out $\eta_{{\rm LO}} = 3.7\%$, $\eta_{{\rm QCD}}^{{\rm I}} = 12.9\%$ and $\eta_{{\rm NLO}}^{{\rm I}} = 12.1\%$ for the integrated cross sections at the LO, QCD NLO and QCD + EW NLO in the inclusive event selection scheme, respectively. We can see that the scale uncertainty at the LO is much less than at the NLO, since the strong coupling constant $\alpha_s$ only appears in the NLO matrix elements. As for the exclusive event selection scheme, we get $\eta_{{\rm QCD}}^{{\rm II}} =2.4\%$ and $\eta_{{\rm NLO}}^{{\rm II}} = 3.3\%$ at the QCD NLO and QCD+EW NLO, respectively. The scale uncertainties are heavily suppressed in the exclusive event selection scheme, therefore we can conclude that the scale uncertainty at the QCD + EW NLO mainly comes from the real emission contributions in the QCD correction, and can be reduced by applying a jet veto. However, in the jet veto scheme (i.e., the scheme II) the jet transverse momentum cut would induce large logarithms and therefore introduces an additional source of theoretical uncertainty. Thus, the resummation should be included in further calculations.
\begin{table}[ht!]
\center\small
\begin{tabular}{cccccccc}
\hline
\hline
\\ [-1.7ex]
{$\mu$}  &  {Scheme}   & ~{$\sigma_{{\rm LO}}$~(fb)}~ & ~{$\Delta \sigma_{{\rm QCD}}$~(fb)}~ & ~{$\Delta \sigma_{{\rm QED}}$~(fb)}~ & ~{$\Delta \sigma_{{\rm EW}}$~(fb)}~ &  ~{$\sigma_{{\rm QCD}}$~(fb)}~  &  ~{$\sigma_{{\rm NLO}}$~(fb)}~  \\
\\ [-1.7ex]
\hline
\\ [-1.7ex]
\multirow{2}{*}{$\frac{1}{4} \mu_0$}  & I & 8.725(2) & 5.257(2) & -0.1524(1) & -1.0692(4) & 15.937(3) & 14.342(4)  \\
&  II & 8.725(2) &  0.374(3) & -0.1543(1) & -1.0711(4) & 11.054(3) &  9.945(4)  \\
\\ [-1.7ex]
\\ [-1.7ex]
\multirow{2}{*}{$\frac{1}{2} \mu_0$}  & I & 8.897(2) & 4.417(2) & -0.1473(1) & -1.0584(4) & 15.216(3) & 13.724(4)  \\
&  II & 8.897(2) & 0.280(2) & -0.1493(1) & -1.0604(4) &  11.078(3)  &  9.990(4)  \\
\\ [-1.7ex]
\\ [-1.7ex]
\multirow{2}{*}{$\mu_0$}  & I & 9.002(2) & 3.860(2) & -0.1403(1) & -1.0401(3) &  14.688(3) & 13.277(4)  \\
&  II & 9.002(2) & 0.313(2) & -0.1425(1) & -1.0423(3) & 11.140(3) &  10.068(3)  \\
\\ [-1.7ex]
\\ [-1.7ex]
\multirow{2}{*}{$2\mu_0$}  & I & 9.053(2) & 3.521(2) & -0.1320(1) & -1.0186(3) & 14.318(3) & 12.967(4)  \\
&  II & 9.053(2) & 0.434(2) & -0.1344(1) & -1.0210(3) & 11.231(3) &  10.169(3)  \\
\\ [-1.7ex]
\\ [-1.7ex]
\multirow{2}{*}{$4\mu_0$} & I & 9.060(2) & 3.320(2) & -0.1230(1) & -0.9934(3) &  14.037(3) & 12.736(3) \\
& II & 9.060(2) & 0.610(2) & -0.1255(1) & -0.9958(3) &  11.327(3)  &  10.274(3)  \\
\\ [-1.7ex]
\hline
\hline
\end{tabular}
\caption{\small The scale dependence of the LO, NLO QCD and NLO QCD + EW corrected integrated cross sections and the corresponding corrections for the $ZZZ$ production at the $\sqrt{s}= 13~{\rm TeV}$ LHC in scheme I and II. }
\label{tab:mu-dependence}
\end{table}

\par
\subsection{Kinematic distributions}
\label{sec:kinedistri}
\par
Now we investigate the kinematic distributions of the final $Z$-bosons as well as the subsequent $Z$-boson leptonic decay products $(Z \to \ell^{+}\ell^{-},~ \ell = e, \mu, \tau)$ for the $ZZZ$ production at the $13~{\rm TeV}$ LHC by taking $\mu_R = \mu_F = \mu_0$. In dealing with the subsequent $Z$-boson decays, we transform our event records into Les Houches event files \cite{lhe} so that we could take use of the {\sc MadSpin} method \cite{madspin:theory, madspin:program} to obtain the $Z$-boson decayed events, which contain the off-shell contributions and spin correlations from the corresponding decays.

\par
\begin{figure}[ht!]
\centering
\includegraphics[scale=0.31]{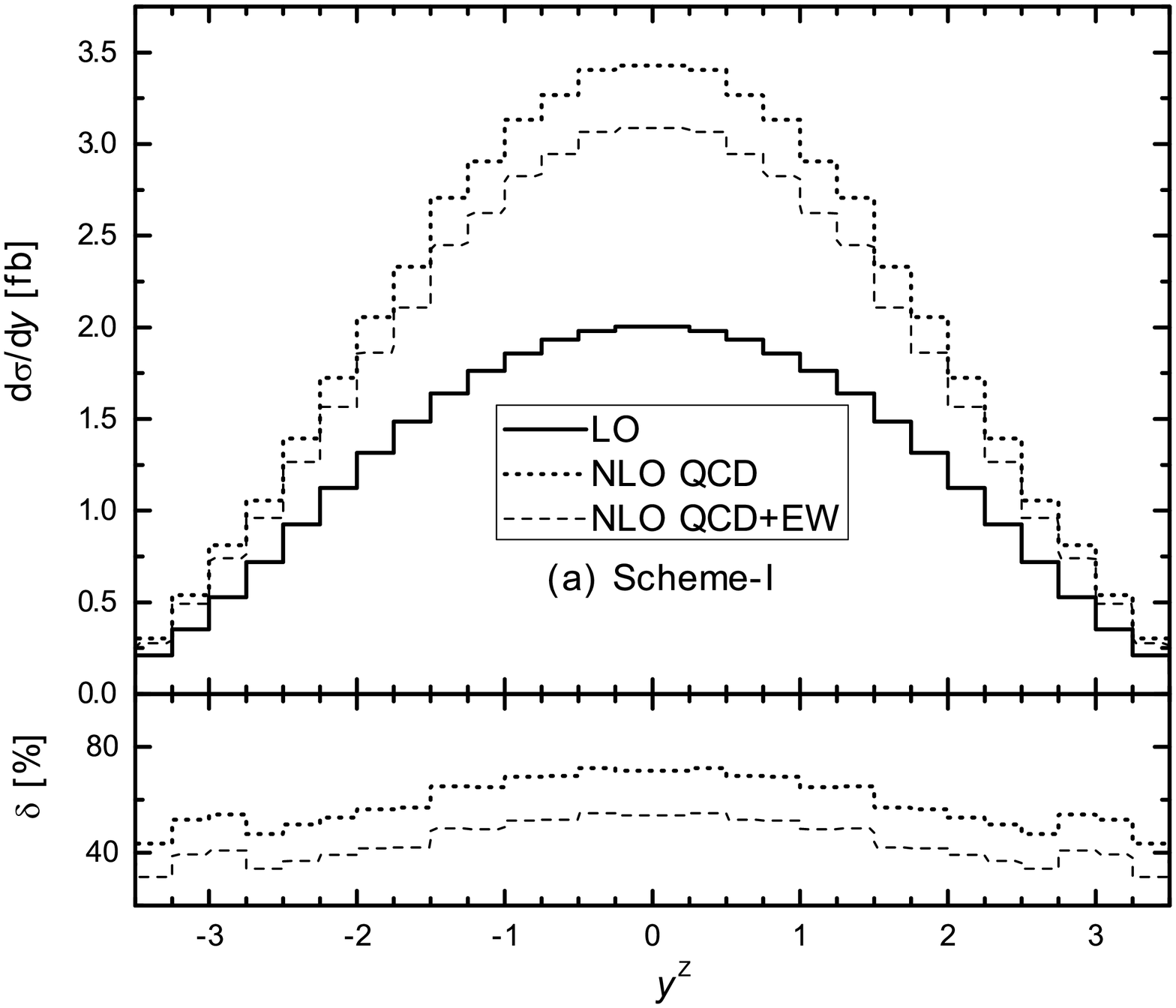}
\includegraphics[scale=0.31]{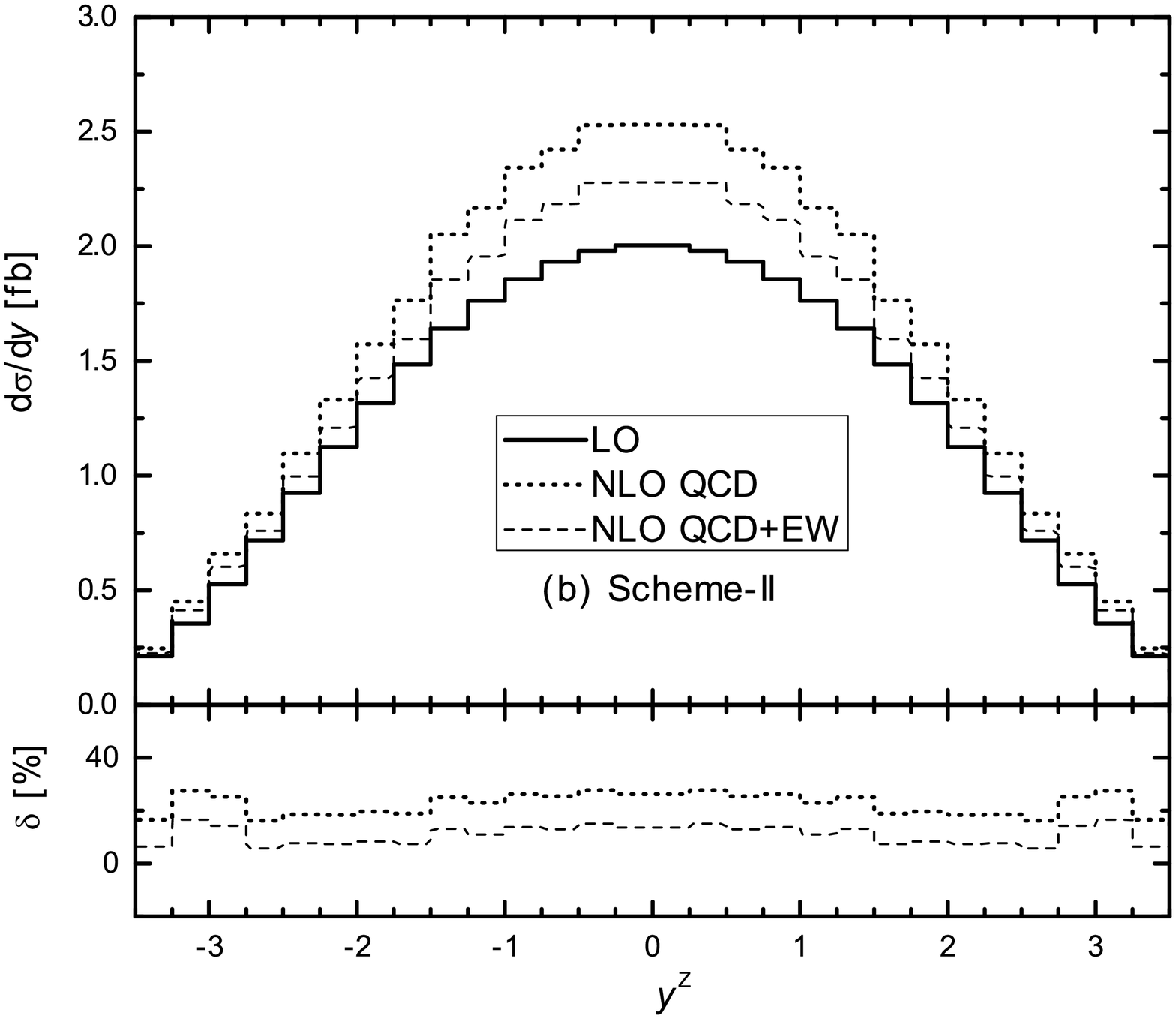}
\caption{\small The LO (solid), NLO QCD (dotted), NLO QCD + EW (dashed) corrected rapidity distributions of the final $Z$-bosons, and the corresponding relative corrections for the $pp \to ZZZ + X$ process at the
$\sqrt{s} = 13~{\rm TeV}$ LHC in the (a) inclusive and (b) exclusive event selection schemes. } \label{fig:zy}
\end{figure}

We provide the LO, NLO QCD, NLO QCD+EW corrected rapidity distributions of the final $Z$-bosons, and the corresponding relative corrections for the $pp \to ZZZ + X$ process at the $\sqrt{s} = 13~{\rm TeV}$ LHC by adopting two event selection schemes in Figs.\ref{fig:zy}(a) and (b) separately. All the rapidities of the three identical $Z$-bosons are filled into the histograms, thus the final differential cross sections should be normalized by multiplying $\frac{1}{3}$\footnote{All the kinematic distributions of the final identical $Z$-bosons and the subsequent leptonic decay products are defined in this way.}. From the figures, we can see that the spacing between $\delta_{{\rm QCD}}$ and $\delta_{{\rm NLO}}$ is relatively fixed in the plotted rapidity region in both of the two event selection schemes. This implies that the NLO EW relative correction, being about $-9.6\%$, hardly depends on the $Z$-boson rapidity. In addition, the NLO QCD relative corrections reach their maxima of about $71\%$ and $26.2\%$ at $y^{Z} = 0$ in the inclusive and exclusive event selection schemes, respectively. Compared to that in the inclusive event selection scheme, the NLO QCD relative correction in the exclusive event selection scheme is heavily reduced. For an event with small $Z$-boson rapidity, one of the final $Z$-bosons would have a large transverse momentum; therefore, the final jet prefers to be energetic and the event would be excluded after applying a jet veto. That is why the NLO QCD relative correction in the exclusive event selection scheme is much smaller than in the inclusive event selection scheme in the central rapidity region.

\par
\begin{figure}[ht!]
\centering
\includegraphics[scale=0.31]{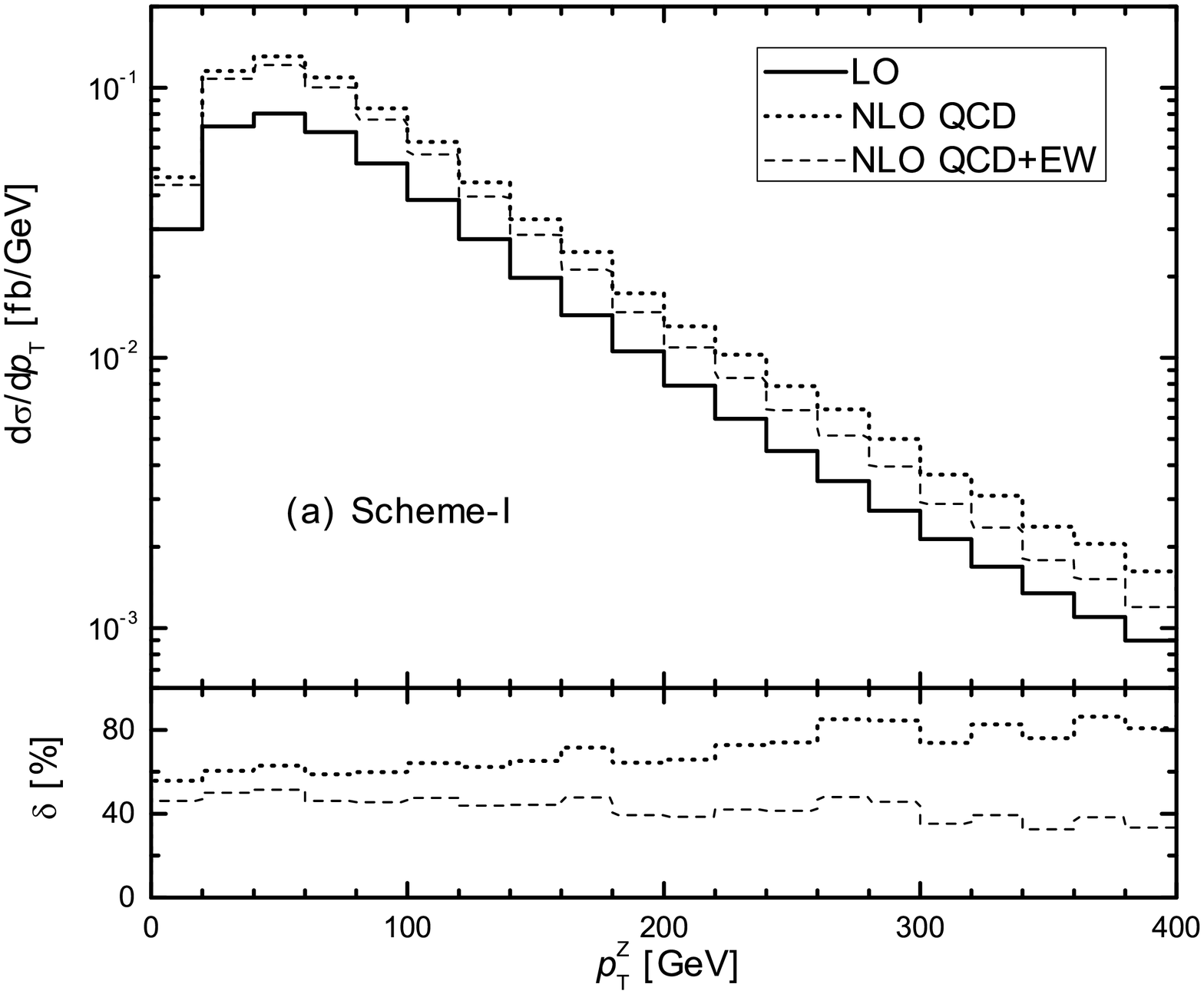}
\includegraphics[scale=0.31]{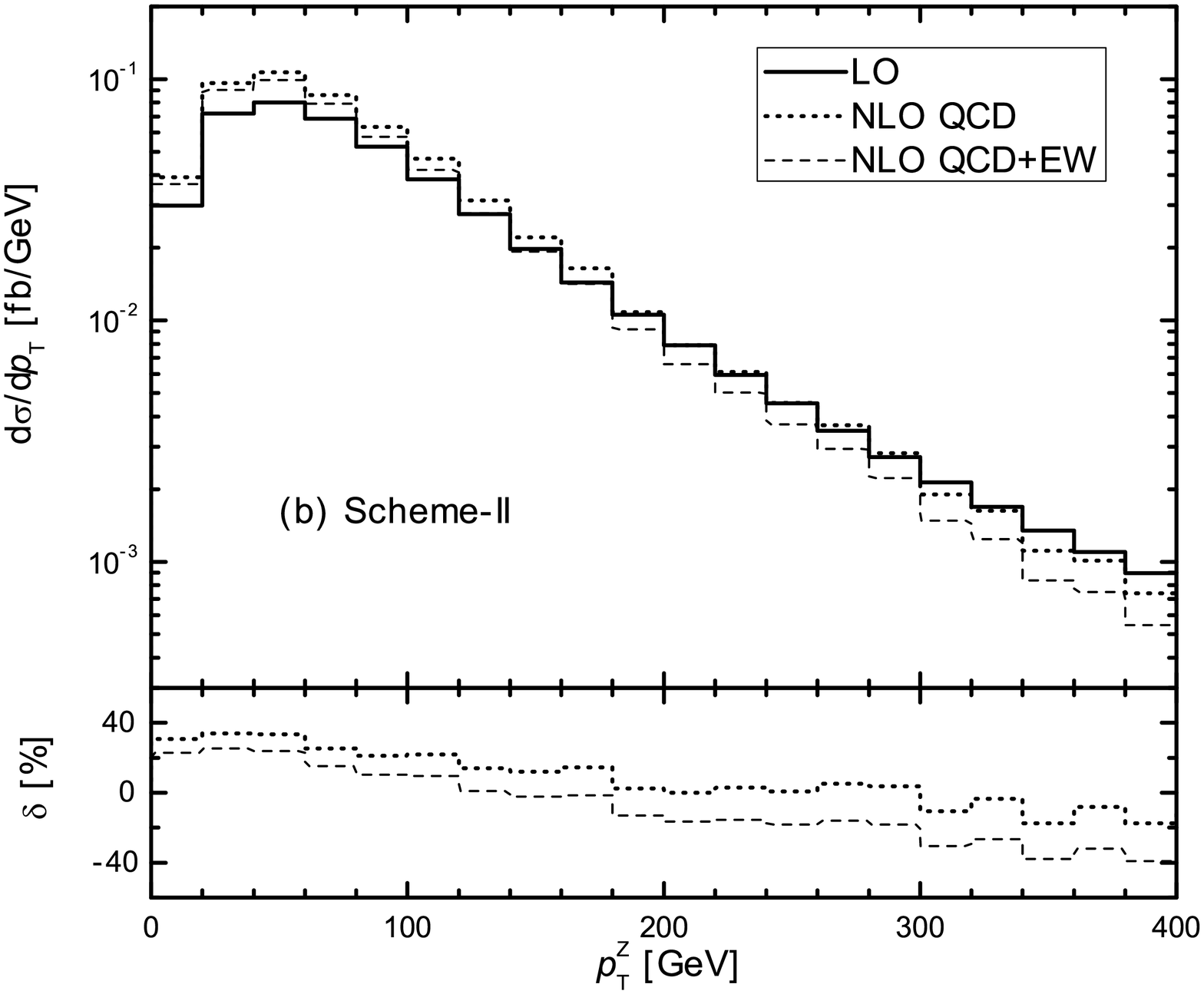}
\caption{\small The  LO (solid), NLO QCD (dotted), NLO QCD + EW (dashed) corrected transverse momentum distributions of the final $Z$-bosons, and the corresponding relative corrections for the $pp \to ZZZ + X$ process at the $\sqrt{s} = 13~{\rm TeV}$ LHC in the (a) inclusive and (b) exclusive event selection schemes. } \label{fig:zpt}
\end{figure}

In Figs.\ref{fig:zpt}(a) and (b) we depict the LO, NLO QCD, NLO QCD + EW corrected transverse momentum distributions of the final $Z$-bosons, and the corresponding relative corrections for the $pp \to ZZZ + X$ process at the $\sqrt{s} = 13~{\rm TeV}$ LHC in the inclusive and exclusive event selection schemes separately. We can see that each distribution reaches its maximum in the vicinity of $p^Z_{{\rm \,T}} \sim 50~{\rm GeV}$. In the inclusive event selection scheme, the NLO QCD relative correction increases to about $80.7\%$ with the increment of $p_{{\rm \,T}}^{Z}$ to $400~ {\rm GeV}$. The NLO EW relative correction is negative in the plotted $p_{{\rm \,T}}^Z$ region. Its absolute value increases significantly in the high $p^Z_{{\rm \,T}}$ region, with an increment of $p_{{\rm \,T}}^Z$ due to the Sudakov effect \cite{Sudakov-1, Sudakov-2}, and reaches about $26.2\%$ at $p_{{\rm \,T}}^Z = 400~ {\rm GeV}$. As a result, the full NLO QCD + EW relative correction gradually reduces from $51.5\%$ to $33.4\%$ with an increment of $p^Z_{{\rm \,T}}$. Similar to the discussion on the $Z$-boson rapidity distribution, for an event with a large $p^Z_{{\rm \,T}}$, the radiated jet tends to carry a large transverse momentum, meaning the event would therefore be discarded in the exclusive event selection scheme. Thus it is reasonable that both $\delta_{{\rm QCD}}$ and $\delta_{{\rm NLO}}$ in the exclusive event selection scheme decrease with an increment of $p^Z_{{\rm \,T}}$, and fall to about $-17.4\%$ and $-39.1\%$ at $p^Z_{{\rm \,T}} = 400~{\rm GeV}$, respectively.

\par
We study the kinematic distributions of the final leptons for the $pp \to ZZZ \to \ell^+_1 \ell^-_1 \ell^+_2 \ell^-_2 \ell^+_3 \ell^-_3 + X$ process in the following discussion. Due to the CP conservation, the kinematic distributions for $\ell^+$ should be the same as the corresponding ones for $\ell^-$. Therefore, we only plot the distributions of one kind of charged lepton, either $\ell^+$ or $\ell^-$.

\par
\begin{figure}[ht!]
\centering
\includegraphics[scale=0.31]{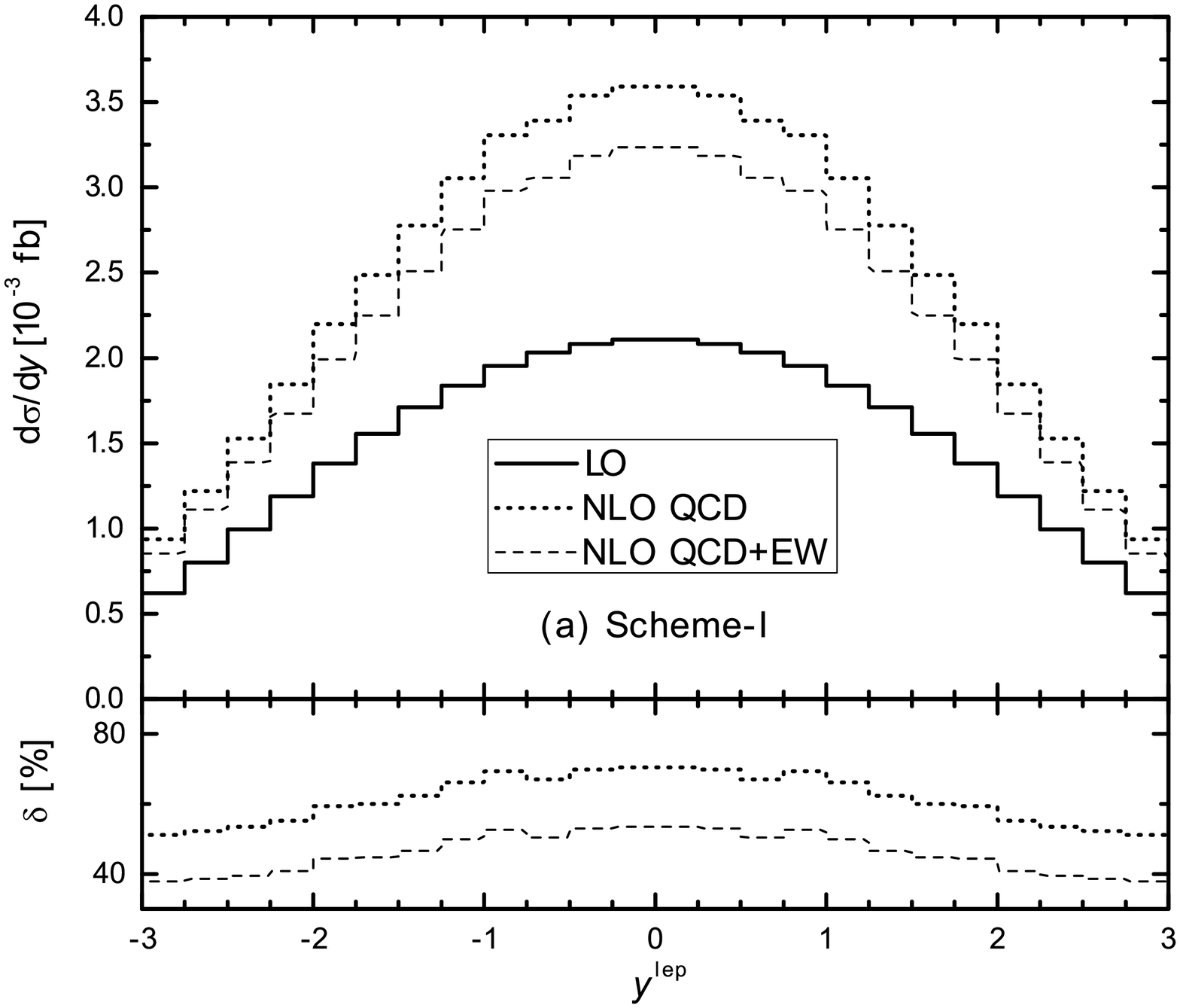}
\includegraphics[scale=0.31]{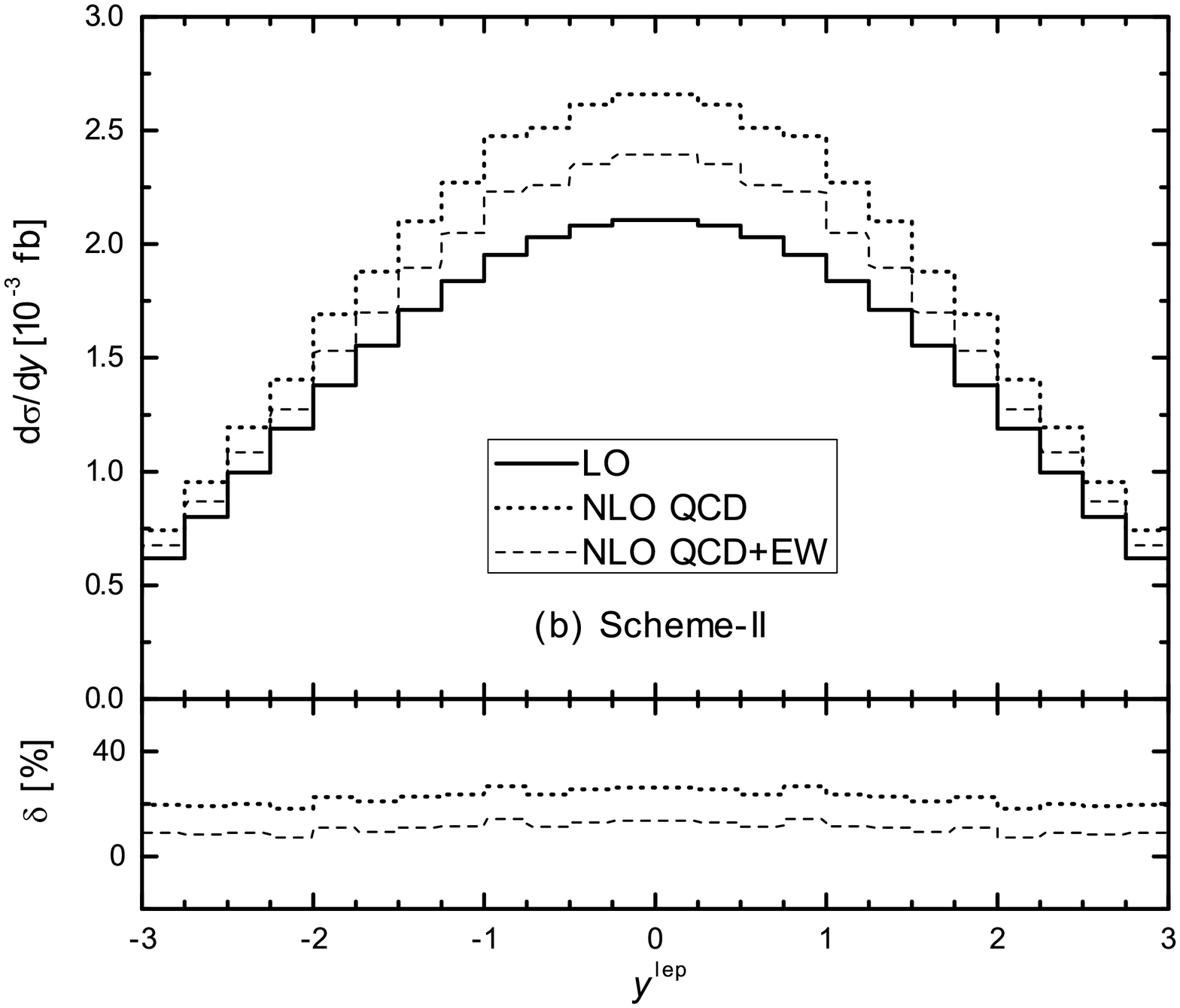}
\caption{\small The LO (solid), NLO QCD (dotted), NLO QCD + EW (dashed) corrected rapidity distributions of the final leptons, and the corresponding relative corrections for the $pp \to ZZZ \to \ell^+_1 \ell^-_1 \ell^+_2 \ell^-_2 \ell^+_3 \ell^-_3 + X$ process at the $\sqrt{s} = 13~{\rm TeV}$ LHC in the (a) inclusive and (b) exclusive event selection schemes. } \label{fig:ly}
\end{figure}
\begin{figure}[ht!]
\centering
\includegraphics[scale=0.31]{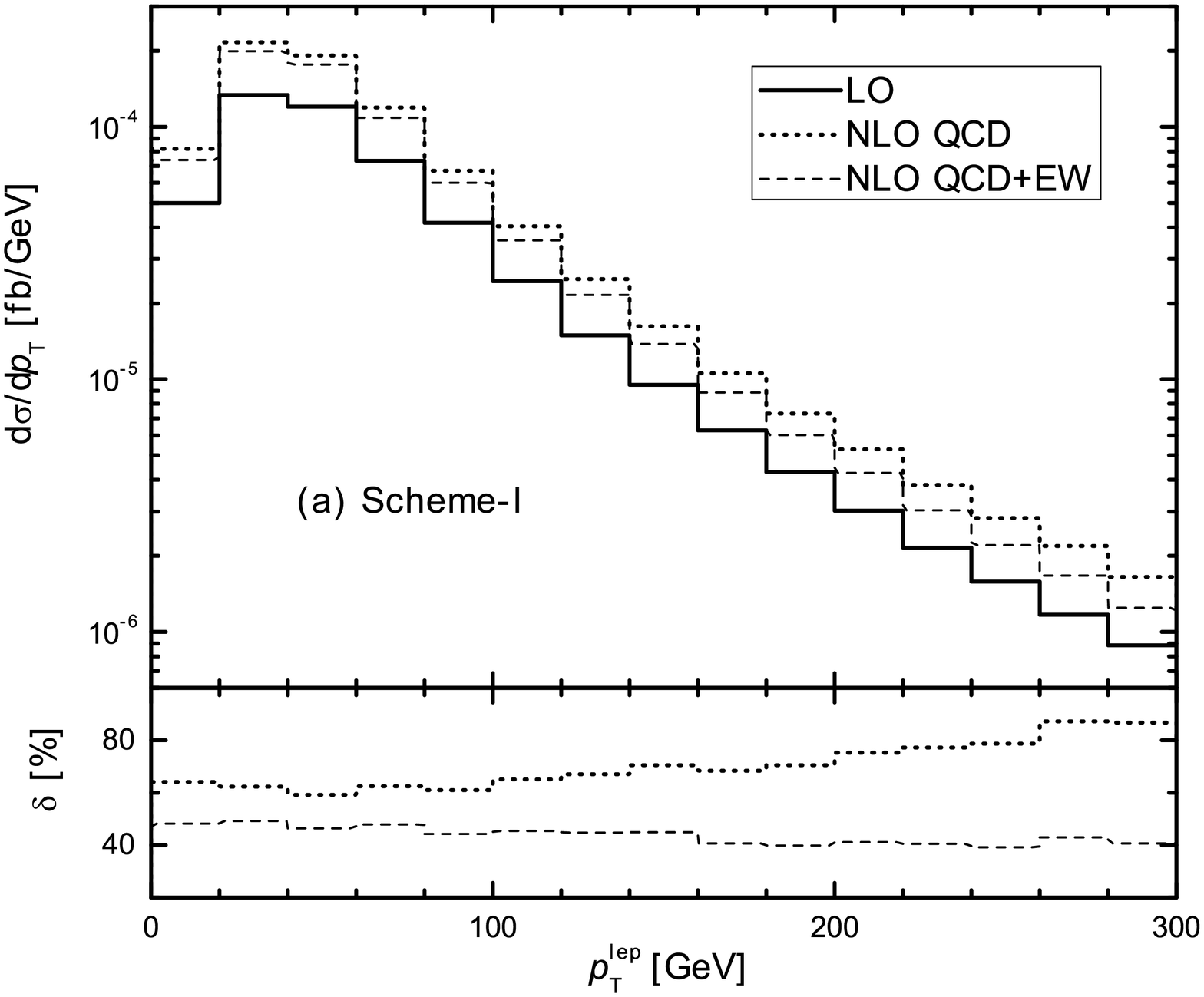}
\includegraphics[scale=0.31]{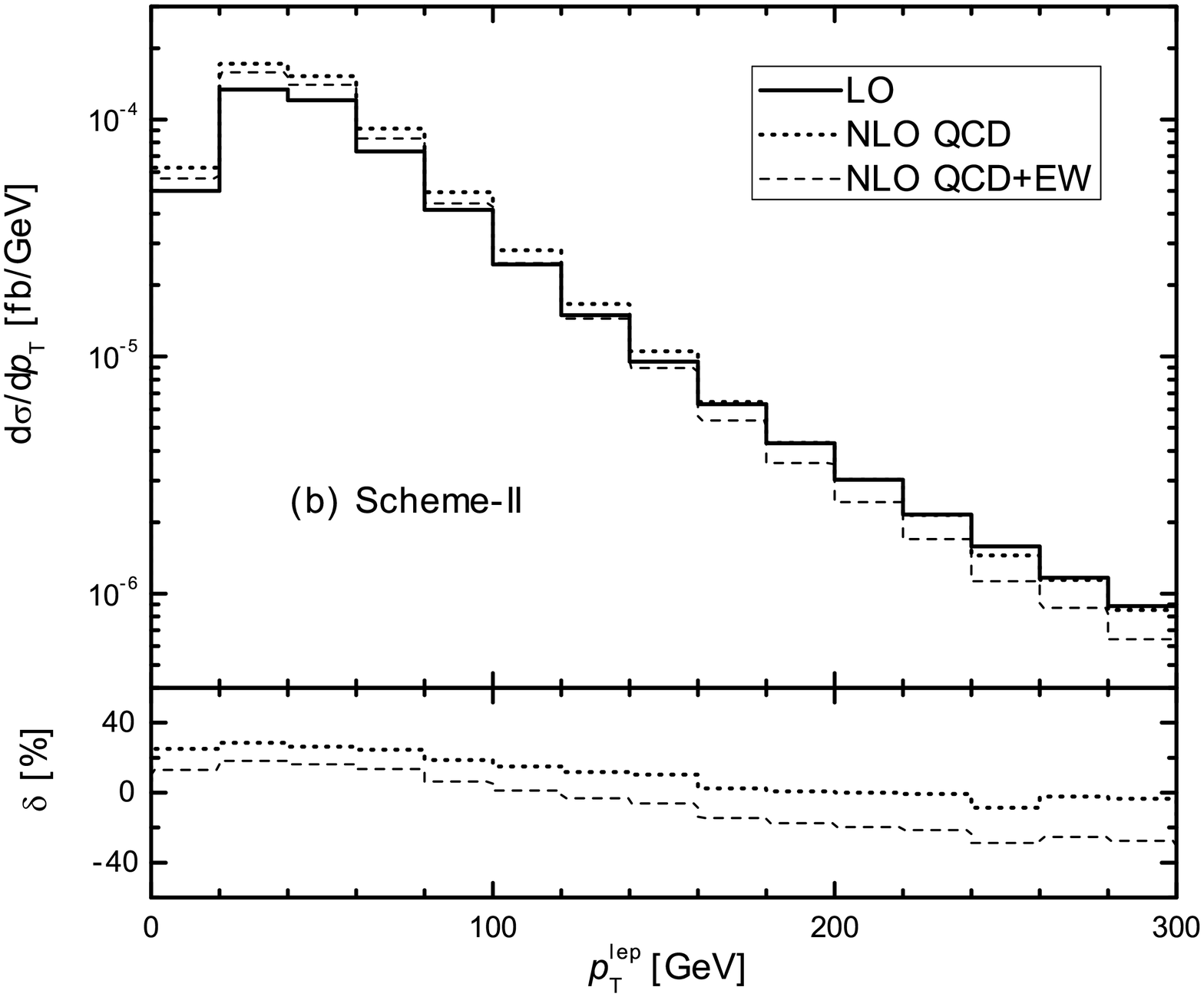}
\caption{\small The LO (solid), NLO QCD (dotted), NLO QCD + EW (dashed) corrected transverse momentum distributions of the final leptons, and the corresponding relative corrections for the $pp \to ZZZ \to \ell^+_1 \ell^-_1 \ell^+_2 \ell^-_2 \ell^+_3 \ell^-_3 + X$ process at the $\sqrt{s} = 13~{\rm TeV}$ LHC in the (a) inclusive and (b) exclusive event selection schemes. } \label{fig:lpt}
\end{figure}

In Figs.\ref{fig:ly}(a), (b) and Figs.\ref{fig:lpt}(a), (b) we present the LO, NLO QCD and NLO QCD + EW corrected rapidity and transverse momentum distributions of the final leptons separately. These figures show that the line shape and variation tendency of each relative correction distribution exhibit similar behavior to the corresponding one in the plots for the $Z$-bosons. As depicted in Figs.\ref{fig:ly}(a) and (b), $\delta_{{\rm QCD}}$ and $\delta_{{\rm NLO}}$ decrease from $70.5\%$ to $51.2\%$ and from $53.5\%$ to $37.8\%$ in the inclusive event selection scheme, while they decrease from $26.2\%$ to $19.7\%$ and from $13.6\%$ to $9.1\%$ in the exclusive event selection scheme, as an increment of $|y^{{\rm lep}}|$ from $0$ to $3$. In both event selection schemes, the NLO EW relative correction varies in the range of $[-10.0\%, -8.8\%]$.

\par
From Figs.\ref{fig:lpt}(a) and (b) we see that the transverse momentum distributions of the final leptons peak at $p^{{\rm lep}}_{{\rm \,T}} \sim 30~{\rm GeV}$. In the inclusive event selection scheme, the NLO QCD relative correction increases from $62.3\%$ to $86.8\%$, while the NLO QCD + EW relative correction decreases from $49.2\%$ to $40.6\%$, with an increment of $p^{{\rm lep}}_{{\rm \,T}}$ from $30$ to $300~{\rm GeV}$. In the exclusive event selection scheme, $\delta_{{\rm QCD}}$ and $\delta_{{\rm NLO}}$ become smaller when $p^{{\rm lep}}_{{\rm \,T}}$ increases, and vary from $28.6\%$ to $-3.5\%$ and from $18.2\%$ to $-27.4\%$, respectively.

\vskip 5mm
\section{Summary}
\label{sec:conclusions}
\par
Precision theoretical predictions for triple gauge boson productions at hadron colliders are important for investigating quartic gauge boson couplings and the electroweak symmetry breaking mechanism. In this work, we study the NLO QCD + NLO EW corrections to the $ZZZ$ production followed by the $Z$-boson leptonic decays at hadron colliders. Our results show that both the NLO QCD and NLO EW corrections are significant, and the NLO EW relative corrections to the integrated cross sections at $\sqrt{s}=13$, $14$, $33$, $100~{\rm TeV}$ hadron colliders are all about $-10\%$. We discuss the dependence of the integrated cross section on the factorization/renormalization scale and the event selection scheme. We also find that the involvement of the strong coupling constant in the NLO QCD correction would contaminate the tiny scale dependence of the LO prediction, but it could be refined by adopting the jet veto scheme. We present some important kinematic distributions of $Z$-bosons as well as leptonic decay products by adopting the {\sc MadSpin} method. This shows that the NLO EW relative correction becomes remarkable in the high energy region due to the EW Sudakov effect.

\par
So far, all the productions of $VV'V''~(V,V',V''=W~ {\rm or}~ Z$) at the LHC have been calculated up to the QCD + EW NLO. The investigations show that the NLO QCD corrections to these processes in the inclusive event selection scheme are remarkable, and the relative corrections are at a level of dozens of percent. We also find that the NLO QCD correction can be heavily suppressed by applying a jet veto cut, which would introduces an additional source of theoretical uncertainty. The NLO EW corrections to the triple weak gauge boson productions are normally negative, and are relatively small compared to the corresponding QCD ones. In precision theoretical studies, these NLO EW corrections should be considered together with the NLO QCD corrections.

\vskip 5mm
\section*{Acknowledgments}
This work was supported in part by the National Natural Science
Foundation of China (No.11275190, No.11375008, No.11375171,
No.11405173, No.11535002).


\vskip 5mm
\begin{thebibliography}{99}

\bibitem{Higgs-ATLAS}
ATLAS collaboration, G. Aad et al., {\it Observation of a new particle in the search for the Standard Model Higgs boson with the ATLAS detector at the LHC}, Phys. Lett. {\bf B 716} (2012) 1 [arXiv:1207.7214].

\bibitem{Higgs-CMS}
CMS collaboration, S. Chatrchyan et al., {\it Observation of a new boson at a mass of 125 GeV with the CMS experiment at the LHC}, Phys. Lett. {\bf B 716} (2012) 30 [arXiv:1207.7235].

\bibitem{qgbc}
S. Godfrey, {\it Quartic gauge boson couplings}, AIP Conf. Proc. {\bf 350} (1995) 209 [arXiv:hep-ph/9505252].

\bibitem{ewsb}
O.J.P. \'Eboli, M.C. Gonzalez-Garcia and S.M. Lietti, {\it Bosonic quartic couplings at CERN LHC}, Phys. Rev. {\bf D 69} (2004) 095005 [arXiv:hep-ph/0310141].

\bibitem{WWZ}
V. Hankele and D. Zeppenfeld, {\it QCD corrections to hadronic $WWZ$ production with leptonic decays}, Phys. Lett. {\bf B 661} (2008) 103 [arXiv:0712.3544].

\bibitem{ZZZ}
A. Lazopoulos, K. Melnikov and F. Petriello, {\it QCD corrections to triboson production}, Phys. Rev. {\bf D 76} (2007) 014001 [arXiv:hep-ph/0703273].

\bibitem{WWW-WZZ}
F. Campanario, V. Hankele, C. Oleari, S. Prestel and D. Zeppenfeld, {\it QCD corrections to charged triple vector boson production with leptonic decay}, Phys. Rev. {\bf D 78} (2008) 094012 [arXiv:0809.0790].

\bibitem{WWZ+ZZZ+WWW+WZZ}
T. Binoth, G. Ossola, C.G. Papadopoulos and R. Pittau, {\it NLO QCD corrections to tri-boson production}, JHEP {\bf 06} (2008) 082 [arXiv:hep-ph/0804.0350]

\bibitem{WWr-ZZr}
G. Bozzi, F. Campanario, V. Hankele and D. Zeppenfeld, {\it Next-to-leading order QCD corrections to $W^+W^-\gamma$ and $ZZ\gamma$ production with leptonic decays}, Phys. Rev. {\bf D 81} (2010) 094030 [arXiv:0911.0438].

\bibitem{Zrr-rrr}
G. Bozzi, F. Campanario, M. Rauch and D. Zeppenfeld, {\it $Z\gamma\gamma$ production with leptonic decays and triple photon production at next-to-leading order QCD}, Phys. Rev. {\bf D 84} (2011) 074028 [arXiv:1107.3149].

\bibitem{Wrr}
U. Baur, D. Wackeroth and M.M. Weber, {\it Radiative corrections to $W\gamma\gamma$ production at the LHC}, PoS RADCOR2009 {\bf 067} (2010) [arXiv:1001.2688].

\bibitem{Wrr-decay}
G. Bozzi, F. Campanario, M. Rauch and D. Zeppenfeld, {\it $W\gamma\gamma$ production with leptonic decays at next-to-leading order QCD}, Phys. Rev. {\bf D 83} (2011) 114035 [arXiv:1103.4613].

\bibitem{WZr}
G. Bozzi, F. Campanario, M. Rauch, H. Rzehak and D. Zeppenfeld, {\it NLO QCD corrections to $W^{\pm}Z\gamma$ production with leptonic decays}, Phys. Lett. {\bf B 696} (2011) 380 [arXiv:1011.2206].

\bibitem{EWWWZ}
D.T. Nhung, L.D. Ninh and M.M. Weber, {\it NLO corrections to $WWZ$ production at the LHC}, JHEP {\bf 12} (2013) 096 [arXiv:1307.7403].

\bibitem{EWWZZ}
Shen Yong-Bai, Zhang Ren-You, Ma Wen-Gan, Li Xiao-Zhou, Zhang Yu, Guo Lei, {\it NLO QCD + NLO EW corrections to WZZ productions with leptonic decays at the LHC}, JHEP {\bf 10} (2015) 186 [arXiv:1507.03693].

\bibitem{EWWWW}
Shen Yong-Bai, Zhang Ren-You, Ma Wen-Gan, Li Xiao-Zhou, Guo Lei, {\it NLO QCD + EW corrections to WWW production with leptonic decays at the LHC} [arXiv:1605.00554].

\bibitem{wishlist-2013}
J. Butterworth et al., {\it Les Houches 2013: Physics at TeV Colliders: Standard Model Working Group Report}, [arXiv:1405.1067].

\bibitem{wishlist-2015}
S. Badger, et al., {\it Les Houches 2015: Physics at TeV Colliders: Standard Model Working Group Report}, [arXiv:1605.04692].

\bibitem{feynarts}
T. Hahn, {\it Generating Feynman diagrams and amplitudes with FeynArts 3}, Comput. Phys. Commun. {\bf 140} (2001) 418 [arXiv:hep-ph/0012260].

\bibitem{formcalc}
T. Hahn and M. P\'erez-Victoria, {\it Automated one-loop calculations in four and $D$ dimensions}, Comput. Phys. Commun. {\bf 118} (1999) 153 [arXiv:hep-ph/9807565].

\bibitem{ff}
G.J. van Oldenborgh, {\it FF - a package to evaluate one-loop Feynman diagrams}, Comput. Phys. Commun. {\bf 66} (1991) 1.

\bibitem{gram-fb}
Fawzi Boudjema, Le Duc Ninh, Sun Hao, Marcus M. Weber, {\it NLO corrections to $e^+e^- \to WWZ$ and $e^+e^- \to ZZZ$}, Phys. Rev. {\bf D 81} (2010) 073007, [arXiv:hep-ph/0912.4234].

\bibitem{gram-cc}
Chen Chong, Ma Wen-Gan, Zhang Ren-You, Zhang Yu, Chen Liang-Wen, Guo Lei, {\it Electroweak radiative corrections to $W^{+}W^{-}\gamma$ production at the ILC}, Eur. Phys. J. {\bf C74}, 3166 (2014), [arXiv:hep-ph/1409.4900].

\bibitem{madgraph}
J. Alwall, R. Frederix, S. Frixione, V. Hirschi, F. Maltoni, O. Mattelaer, H.-S. Shao, T. Stelzer, P. Torrielli, M. Zaro, {\it The automated computation of tree-level and next-to-leading order differential cross sections, and their matching to parton shower simulations}, JHEP {\bf 07} (2014) 079 [arXiv:hep-ph/1405.0301].

\bibitem{Denner:EWphysics}
A. Denner, {\it Techniques for the calculation of electroweak radiative corrections at the one-loop level and results for W-physics at LEP200}, Fortschr. Phys. {\bf 41} (1993) 307 [arXiv:0709.1075].

\bibitem{w-production}
S. Dittmaier and M. Kramer, {\it Electroweak radiative corrections to $W$-boson production at hadron colliders}, Phys. Rev. {\bf D 65} (2002) 073007 [arXiv:hep-ph/0109062].

\bibitem{delt-r}
A. Sirlin, {\it Radiative corrections in the $SU(2)_L \times U(1)$ theory: A simple renormalization framework}  Phys. Rev. D {\bf 22}, 971 (1980).

\bibitem{TCPSS}
B.W. Harris and J.F. Owens, {\it Two cutoff phase space slicing method}, Phys. Rev. {\bf D 65} (2002) 094032 [arXiv:hep-ph/0102128].

\bibitem{DIPOLE-1996}
S. Catani and M.H. Seymour, {\it The dipole formalism for the calculation of QCD jet cross sections at Next-to-Leading order}, Phys. Lett. {\bf B 378} (1996) 287 [arXiv:hep-ph/9602277].

\bibitem{DIPOLE-1997}
S. Catani and M. H. Seymour, {\it A general algorithm for calculating jet cross sections in NLO QCD} Nucl. Phys. {\bf B 485} (1997) 291; Erratum-ibid. {\bf B 510} (1998) 503 [arXiv:hep-ph/9605323].

\bibitem{DIPOLE-2000}
S. Dittmaier, {\it A general approach to photon radiation off fermions} Nucl. Phys. {\bf B 565} (2000) 69, [arXiv:hep-ph/9904440].

\bibitem{DIPOLE-alpha-1}
Zoltan Nagy, Zoltan Trocsanyi, {\it Next-to-leading order calculation of four-jet observables in electron-positron annihilation} Phys. Rev. {\bf D 59} (1999) 014020; Erratum-ibid. {\bf D 62} (2000) 099902 [arXiv:hep-ph/9806317].

\bibitem{DIPOLE-alpha-2}
Zoltan Nagy, {\it Next-to-leading order calculation of three-jet observables in hadron-hadron collision} Phys. Rev. {\bf D 68} (2003) 094002 [arXiv:hep-ph/0307268].

\bibitem{DIS-facscheme}
K.-P. O. Diener, S. Dittmaier, and W. Hollik, {\it Electroweak higher-order effects and theoretical uncertainties in deep-inelastic neutrino scattering}, Phys. Rev. {\bf D 72} (2005) 093002 [arXiv:hep-ph/0509084].

\bibitem{FSR}
A.B. Arbuzov, R.R. Sadykov, and Z. Was, {\it QED bremsstrahlung in decays of electroweak bosons}, Eur. Phys. J. {\bf C73}, 2625 (2013) [arXiv:1212.6783].

\bibitem{dress-lepton}
ATLAS collaboration, G. Aad et al., {\it Measurement of the transverse momentum distribution of $Z/\gamma^*$ bosons in proton-proton collisions at $\sqrt{s} = 7$ {\rm TeV} with the ATLAS detector}, Phys. Lett. {\bf B 705} (2011) 415 [arXiv:1107.2381].

\bibitem{PDG}
{\sc Particle Data Group} collaboration, K.A. Olive et al., {\it Review of particle physics}, Chin. Phys. {\bf C 38} (2014) 090001.

\bibitem{NNPDF}
NNPDF collaboration, R.D. Ball et al., {\it Parton distributions with QED corrections}, Nucl. Phys. {\bf B 877} (2013) 290 [arXiv:1308.0598].

\bibitem{Denner-delta}
A. Denner, S. Dittmaier, M. Hecht and C. Pasold, {\it NLO QCD and electroweak corrections to $W+\gamma$ production with leptonic $W$-boson decays}, JHEP {\bf 04} (2015) 018 [arXiv:1412.7421].

\bibitem{lhe}
J. Alwall et al., {\it A standard format for Les Houches Event Files}, Comput. Phys. Commun. {\bf 176} (2007) 300 [arXiv:hep-ph/0609017].

\bibitem{madspin:theory}
S. Frixione, E. Laenen, P. Motylinski, and B.R. Webber, {\it Angular correlations of lepton pairs from vector boson and top quark decays in Monte Carlo simulations} JHEP {\bf 04} (2007) 081 [arXiv:hep-ph/0702198].

\bibitem{madspin:program}
P. Artoisenet, R. Frederix, O. Mattelaer, and R. Rietkerk, {\it Automatic spin-entangled decays of heavy resonances in Monte Carlo simulations} JHEP {\bf 03} (2013) 015  [arXiv:1212.3460].

\bibitem{Sudakov-1}
V. V. Sudakov, Sov. Phys. JETP {\bf 3} (1956) 65.

\bibitem{Sudakov-2}
V. S. Fadin, L. N. Lipatov, A. D. Martin and M. Melles, {\it Resummation of double logarithms in electroweak high energy processes} Phys. Rev. {\bf D 61} (2000) 094002 [arXiv:hep-ph/9910338].

\end{thebibliography}
\end{document}